\documentclass[aoas,nameyear,dvips]{arximspdf}
\usepackage{graphicx}
\usepackage{breakurl}

\doi{10.1214/10-AOAS398D}
\referstodoi{10.1214/10-AOAS398}
\volume{5}
\issue{1}
\pubyear{2011}
\firstpage{65}
\lastpage{70}

\begin{document}
\begin{frontmatter}

\title{Discussion of: A statistical analysis of multiple temperature
proxies: Are
reconstructions of surface temperatures over the last\\ 1000 years
reliable?}
\runtitle{Discussion}
\pdftitle{Discussion on A statistical analysis of multiple temperature
proxies: Are reconstructions of surface temperatures over the last 1000
years reliable?
by B. B. McShane and A. J. Wyner}

\begin{aug}
\author[A]{\fnms{Gavin A.} \snm{Schmidt}\ead[label=e2]{Gavin.A.Schmidt@nasa.gov}},
\author[B]{\fnms{Michael E.} \snm{Mann}\corref{}\ead[label=e1]{mann@psu.edu}}
and
\author[C]{\fnms{Scott D.} \snm{Rutherford}}

\runauthor{G. A. Schmidt, M. E. Mann and S. D. Rutherford}

\affiliation{NASA Goddard Institute for Space Studies, Pennsylvania State University
and~Roger Williams University}

\address[A]{G. A. Schmidt\\
NASA Goddard Institute for Space Studies\\
New York, New York\\
USA\\
\printead{e2}} 

\address[B]{M. E. Mann\\
Department of Meteorology and Earth\\
and Environmental Systems Institute\\
Pennsylvania State University\\
University Park, Pennsylvania\\
USA\\
\printead{e1}}

\address[C]{S. D. Rutherford\\
Department of Environmental Science\\
Roger Williams University\\
Bristol, Rhode Island\\
USA}
\end{aug}

\received{\smonth{9} \syear{2010}}



\end{frontmatter}

McShane and Wyner (\citeyear{MW2011}) (henceforth MW) analyze a dataset of
``proxy'' climate records previously used by Mann et al. (\citeyear{Metal2008})
(henceforth M08) to attempt to assess their utility in reconstructing
past temperatures. MW introduce new methods in their analysis, which is
welcome. However, the absence of both proper data quality control and
appropriate ``pseudoproxy'' tests to assess the performance of
their methods invalidate their main conclusions.

\begin{figure}
\begin{tabular}{@{}c@{}}

\includegraphics{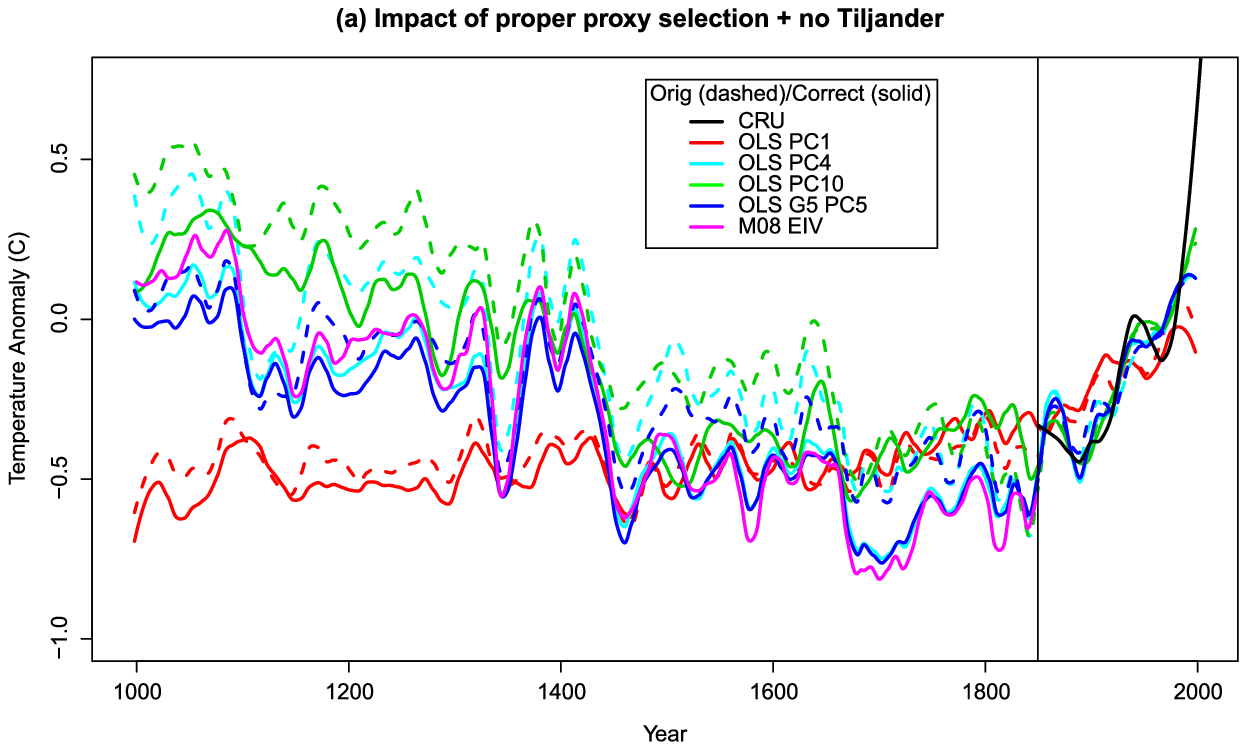}
\\

\includegraphics{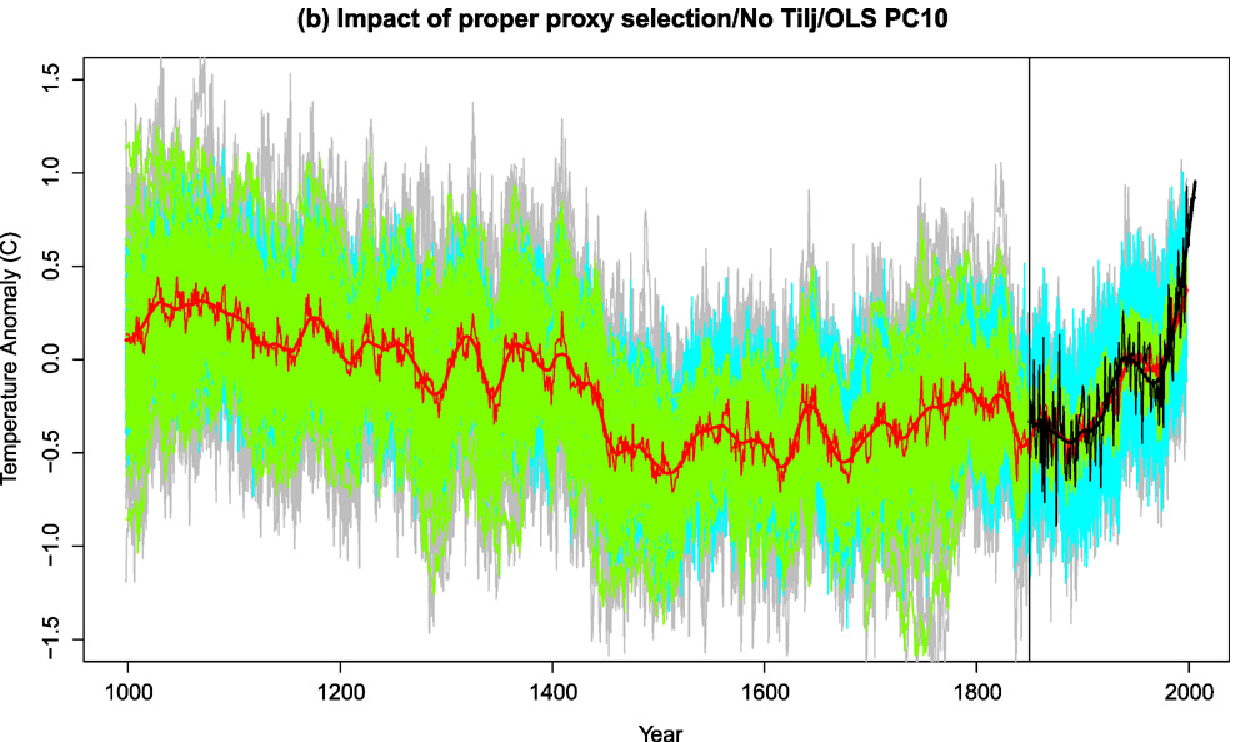}

\end{tabular}
  \caption{Reconstructions of mean Northern Hemisphere land
temperatures over the past millennium for various methodological choices
(cf. MW Figure 14). \textup{(a)} Results using the M08 frozen AD 1000 network of
59 minus 4 ``Tiljander'' proxy records (corresponding results based on
all 59 records are shown in Supplementary Figure S1). Shown for
comparison are the original MW results and the Mann et al. (\protect\citeyear{Metal2008}) ``EIV''
decadal ``CRU'' NH land temperature reconstruction based on the
identical proxy data. The OLS reconstructions have been filtered with a
loess smoother ($\mathrm{span} =0.05$) to emphasize low-frequency (greater than 50
year) variations. Associated annual reconstructions are shown in
Supplementary Figure S2. \textup{(b)} Comparison of Monte Carlo ensemble (and
mean) reconstructions using ``OLS PC10'' as in MW Figure 16. Labeled
reconstructions are in color, grey lines are the total set of MW
reconstructions after allowing for uncertainties in the coefficients.}\label{fig1}
\end{figure}

We deal first with the issue of data quality. In the frozen 1000 AD
network of 95 proxy records used by MW, 36 tree-ring records were not
used by M08 due to their failure to meet objective standards of
reliability. These records did not meet the minimal replication
requirement of at least eight independent contributing tree cores (as
described in the Supplemental Information of M08). That requirement
yields a smaller dataset of 59 proxy records back to AD 1000 as clearly
indicated in M08. MW's inclusion of the additional poor-quality proxies
has a material affect on the reconstructions, inflating the level of
peak apparent Medieval warmth, particularly in their featured
``OLS PC10'' ($K=10$ PCs of the proxy data used as predictors of
instrumental mean NH land temperature) reconstruction. The further
elimination of four potentially contaminated ``Tiljander'' proxies
[as tested in M08; M08 also tested the impact of removing tree-ring
data, including controversial long ``Bristlecone pine'' tree-ring
records. Recent work [cf. Salzer et al. (\citeyear{Setal2009})], however, demonstrates
those data to contain a reliable long-term temperature signal], which
yields a set of 55 proxies, further reduces the level of peak Medieval
warmth (Figure \ref{fig1}(a); cf. Figure 14 in MW; see also Supplementary Figures
S1--S2 [Schmidt, Mann and Rutherford (\citeyear{SMR2010a}, \citeyear{SMR2010b})]).

\setcounter{figure}{0}
\begin{figure}
\begin{tabular}{@{}c@{}}

\includegraphics{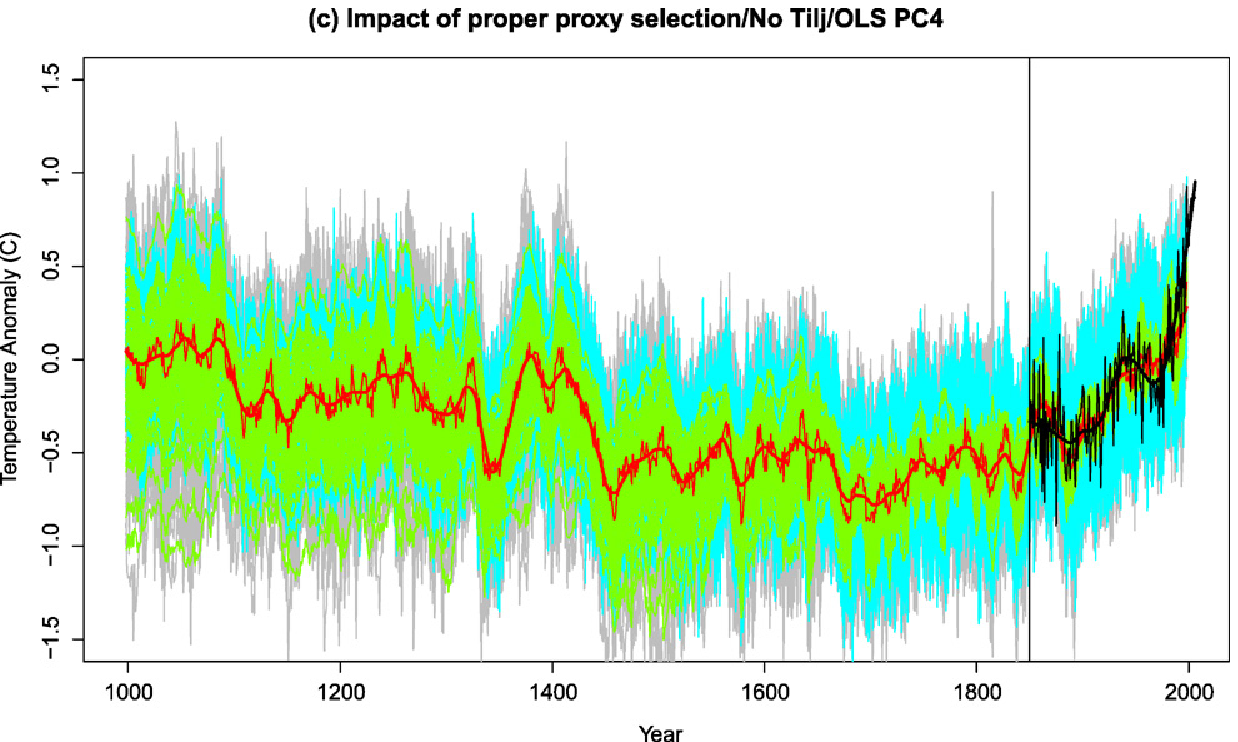}

\end{tabular}
  \caption{\textup{(c)} As in \textup{(b)} above but instead using ``OLS PC4.''}
\end{figure}

The MW ``OLS PC10'' reconstruction has greater peak apparent
\mbox{Medieval} warmth in comparison with M08 or any of a dozen similar
hemispheric temperature reconstructions [Jansen et al. (\citeyear{Jetal2007})]. That
additional warmth, as shown above, largely disappears with the use of
the more \mbox{appropriate} dataset. Using their reconstruction, MW nonetheless
still found recent warmth to be unusual in a long-term context: they
estimate an 80\% probability that the decade 1997--2006 is warmer than
any other for at least the past 1000 years. Using the more appropriate
55-proxy dataset with the same ($K=10$) estimation procedure, we
calculate a higher probability of 86\% that recent decadal
warmth is unprecedented for the past millennium [Figure \ref{fig1}(b)].

However $K=10$ principal components is almost certainly too large, and
the resulting reconstruction likely suffers from statistical
over-fitting. Objective selection criteria applied to the M08 AD 1000
proxy network (see Supplementary Figure S4), as well as independent
``pseudoproxy'' analyses discussed below, favor retaining only
$K=4$ (``OLS PC4'' in the terminology of MW). Using this
reconstruction, we observe a very close match [e.g., Figure \ref{fig1}(a)] with the
relevant M08 reconstruction and we calculate considerably higher
probabilities up to 99\% that recent decadal warmth is unprecedented for
at least the past millennium [Figure~\ref{fig1}(c)]. These posterior probabilities
imply substantially higher confidence than the ``likely''
assessment by M08 and IPCC (\citeyear{IPCC2007}) (a 67\% level of confidence). Indeed,
a probability of 99\% not only exceeds the IPCC ``very likely''
threshold (90\%), but reaches the ``virtually certain'' (99\%)
threshold. However, since these posterior probabilities do not take into
account potential systematic issues in the source data, are sensitive to
methodological choices, and vary by a few percent depending on the MCMC
realization, we maintain that a ``likely'' conclusion is most
consistent with the balance of evidence [IPCC (\citeyear{IPCC2007})].

There are additional methodological weaknesses in the techniques
employed by MW that require discussion. MW mix incommensurate (decadal
vs. annual resolution) proxy data in their procedure, a problem that is
avoided by the ``hybrid'' frequency band calibration method used
by M08. Using a version of the proxy data that was consistently low-pass
filtered to retain only decadal features shows even better agreement
with the M08 reconstruction (supplementary Figure S3).

\begin{figure}

\includegraphics{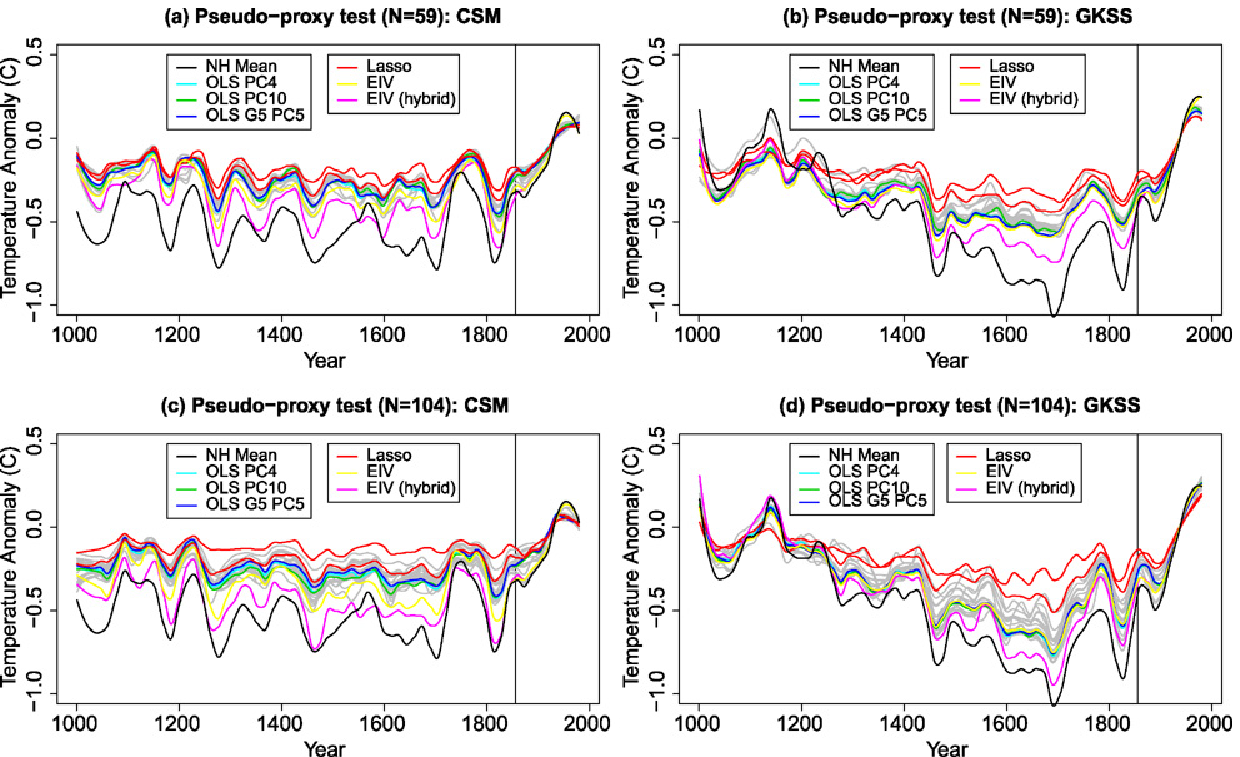}

  \caption{Pseudoproxy tests of reconstruction methodologies
used by MW and comparison with the hybrid and nonhybrid RegEM EIV
methods used by M08. The pseudoproxy networks are defined by a randomly
selected set of gridboxes using two different coupled ocean-atmosphere
general circulation model (OAGCM) simulations subjected to estimated
natural and anthropogenic forcing over the past millennium.
Pseudoproxies are constructed assuming ``red'' proxy noise [$\operatorname{AR}(1)$ with
$\rho =0.32$] yielding mean signal-to-noise amplitude ratio of $\mathrm{SNR}=0.4$,
characteristics which are consistent with estimates from actual proxy
data networks [see Mann et al. (\protect\citeyear{Metal2007})]. All reconstructions use a
calibration interval of 1856--1980. Figure shows results for a
59-location network including \textup{(a)}~NCAR CSM and \textup{(b)} GKSS simulations and
a network with 104 locations for \textup{(c)} CSM and \textup{(d)} GKSS. Labeled
reconstructions are in color, grey lines are the total set of MW
reconstruction techniques. Note that uncertainties are reduced for the
larger network, where the underestimation bias becomes negligible for
the hybrid RegEM EIV method.}\label{fig2}
\end{figure}

Furthermore, methods using simple Ordinary Least Squares (OLS)
regressions of principal components of the proxy network and
instrumental data suffer from known biases, including the
underestimation of variance [see, e.g., Hegerl et al. (\citeyear{Hetal2006})]. The
spectrally ``red'' nature of the noise present in proxy records
poses a particular challenge [e.g., Jones et al. (\citeyear{Jetal2009})]. A standard
benchmark in the field is the use of synthetic proxy data known as
``pseudoproxies'' derived from long-term climate model simulations
where the true climate history is known, and the skill of the particular
method can be evaluated [see, e.g., Mann et al. (\citeyear{Metal2007}); Jones et al.
(\citeyear{Jetal2009}) and numerous references therein]. (We note that the term
``pseudoproxy'' was misused in MW to instead denote various noise
models.) In contrast to the MW claim that their methods perform
``fairly similarly,'' these tests show dramatic differences in
model performance (Figure \ref{fig2}). Indeed, the various flavors of OLS and,
particularly, the ``Lasso'' method (used only in the first half of
MW), suffer from serious underestimation biases in comparison with, for
example, the hybrid RegEM approach of M08 (see also Table S1).

Taken together, these points demonstrate that any conclusions regarding
the utility of proxies in reconstructing past climate drawn by MW were,
at best, overstated. Assessing the skill of methods that do not work
well (such as Lasso) and concluding that no method can therefore work
well, is logically flawed. Additional problems exist in their assessment
procedure---reducing the size of the hold out periods to 30 years from
46 years in M08, for instance, makes it more difficult to meaningfully
diagnose statistical skill.

Problems in climate research, such as statistical climate
reconstruction, require sophisticated statistical approaches and a
thorough understanding of the data used. Moreover, investigations of the
underlying spatial patterns of past climate changes, rather than simply
hemispheric mean temperature estimates, are most likely to provide
insights into climate dynamics [e.g., Mann et al. (\citeyear{Metal2009}), Schmidt
(\citeyear{S2010})]. Further progress in this area will most likely arise from
continuing collaboration between the statistics and climate science
communities, such as fostered since 1996 by the joint NSF/NCAR
Geophysical Statistics Project.

\section*{Acknowledgments}
We thank Sonya Miller for substantial technical assistance. The
JAGS/rjags code used in the Bayesian modeling was adapted from
\url{http://probabilitynotes.wordpress.com/}.

Supplementary figures and tables, data used, and scripts for performing
all analyses are all available at:
\url{http://www.meteo.psu.edu/\textasciitilde mann/supplements/AOAS/}

\begin{supplement}[id=suppA]
\sname{Supplement A}
\stitle{Supplemental figures}
\slink[doi]{10.1214/10-AOAS398DSUPPA}  
\slink[url]{http://lib.stat.cmu.edu/aoas/398D/supplementA.pdf}
\sdatatype{.pdf}
\sdescription{Additional figures S1--4 and Table S1.}
\end{supplement}
\begin{supplement}[id=suppB]
\sname{Supplement B}
\stitle{Code and data for producing all figures and
results in the paper}
\slink[doi]{10.1214/10-AOAS398DSUPPB}  
\slink[url]{http://lib.stat.cmu.edu/aoas/398D/supplementB.zip}
\sdatatype{.zip}
\end{supplement}

\printaddresses

\end{document}